\documentclass{IET}

\usepackage[utf8]{inputenc}
\usepackage[english]{babel}

\usepackage{hyperref}

\usepackage{lipsum}% http://ctan.org/pkg/lipsum
\usepackage{graphicx}% http://ctan.org/pkg/graphicx
\usepackage{lettrine}% big first latter
\usepackage{tikz}
\usetikzlibrary{positioning,calc,arrows}
\usepackage{amsmath}
\usepackage{slashed}
\usepackage{cancel}
\usepackage{float}
\usepackage{lipsum}
\usepackage{stfloats}
\usepackage[bb=boondox]{mathalfa}

{}
{}
{}

\begin{document}

\title{{LDPC Codes over Gaussian Multiple Access Wiretap Channel}}

\author{Sahar~Shahbaz}

\author[1,*]{Bahareh~Akhbari}
\affil{Faculty of Electrical
		Engineering, K. N. Toosi University of Technology, Tehran, Iran.}

\author[2,*]{Reza~Asvadi}
\affil{Faculty of Electrical Engineering, Shahid Beheshti University,
		Tehran, Iran.}

\affil[*]{akhbari@kntu.ac.ir;\,\,r\_asvadi@sbu.ac.ir}

\abstract
{We study the problem of two-user Gaussian multiple access channel (GMAC) in the presence of an external eavesdropper. In this problem, an eavesdropper receives a signal with a lower signal-to-noise ratio (SNR) compared to the legitimate receiver and all transmitted messages should be kept confidential against the eavesdropper. For this purpose, we propose a secure coding scheme on this channel which utilizes low-density parity-check (LDPC) codes by employing random bit insertion and puncturing techniques. At each encoder, the confidential message with some random bits as a random message are systematically encoded, and then the associated bits to the confidential message are punctured. Next, the encoders send their unpunctured bits over a Gaussian multiple access wiretap channel (GMAC-WT). The puncturing distribution applied to the LDPC code is considered in two cases: random and optimized. We utilize a modified extrinsic information transfer (EXIT) chart analysis to optimize the puncturing distribution for each encoder. The \emph{security gap} is used as a measure of secrecy for the sent messages over GMAC-WT which should be made as small as possible.
We compare the achieved secure rate pair with an achievable secrecy rate region of GMAC-WT to show the effective performance of the proposed scheme. In this paper, equal and unequal power conditions at the transmitters are investigated. For both cases, we attain a fairly small security gap which is equivalent to achieve the points near the secrecy rate region of GMAC-WT.
}

\maketitle

\section{Introduction}\label{sec1}

\par Secure communication has been traditionally provided using cryptographic protocols in the upper layers of communication systems. However, some recent security techniques are applied at the physical layer to provide secrecy defined as physical layer secrecy \cite{bloch2011physical}. In these schemes there is no need to use secret/public keys and there is also no need to consider any computational limitations at the eavesdropper.  These are the main advantages of physical layer secrecy schemes in contrast to classical cryptography schemes \cite{bloch2011physical}.
Wyner presented the wiretap channel model in 1975 which consists of one transmitter (Alice), one legitimate receiver (Bob) and one eavesdropper (Eve). He showed that there is a coding scheme which maps a message $M$ to a codeword $X^n$, and if Eve receives a noisy sequence $Z^n$, this coding scheme changes her confusion about $M$ to the maximum possible amount \cite{Wyner}. Thus, the coded message should satisfy the reliability for the legitimate receiver and the confidentiality against the Eve, concurrently. Information-theoretic limits of secret communications for different single user and multiuser channels have recently been investigated \cite{bloch2011physical,6582704}. Moreover, studies show that different families of error correction codes such as LDPC and Polar codes might be used as coding schemes to provide physical layer secrecy \cite{harrison2013coding}. It should be noted that LDPC codes are also used in public key \cite{4809594} or symmetric key cryptosystems \cite{2014.0101} or as a security enhancement technique \cite{5753935}, which are not considered as physical layer security systems, and hence are not at the focus of this paper.

\par Secrecy metrics are typically defined based on the uncertainty of Eve about the sent message conditioned on receiving the noisy sequence $Z^n$, which is called Eve's equivocation \cite{bloch2011physical}. If asymptotic equivocation of Eve is equal to the entropy of the sent message, then a so-called \emph{strong secrecy} condition is established. On the contrary, asymptotic average equivocation of Eve equals the entropy of the sent message in the \emph{weak secrecy} condition. Measuring this equivocation for finite block length codewords over an additive white Gaussian noise channel is difficult \cite{5740591}. To circumvent this problem, researchers provide another criterion. The security gap is defined as the difference in SNRs, in decibel (dB), between Bob's channel with the bit-error rate (BER) close to zero and Eve's channel with the BER close to half \cite{5740591}.

\par In this paper, we consider the problem of Gaussian multiple access wiretap channel (GMAC-WT). The multiple access channel (MAC) consists of two or more sources which send messages to a common receiver. In GMAC-WT, there exists an external eavesdropper that should not access all users' messages \cite{EkremUlukus}. In other words, all messages must be kept secret from Eve. The MAC with confidential messages has been investigated in various scenarios from an information theoretic perspective for which some theoretical bounds have been derived on its secrecy capacity region (see \cite{6582704,EkremUlukus,liang2008multiple,zivari2016imperfect,7762075} and references therein). In this paper, we propose a coding scheme to provide physical layer secrecy over GMAC-WT using LDPC codes which has not been addressed thus far.

\par To provide secrecy through a coding scheme, a set of sub-codebook should exist for each sent message to conceal the secret part of the message \cite{bloch2011physical}. The secure encoder randomly chooses a codeword from the sub-codebook corresponding to the message. Thus, a stochastic encoding is performed from the message point of view. Secure codes should provide a nested structure with random parameters at the encoders. LDPC codes with random bit insertion are one of the code constructions that can satisfy these mentioned conditions \cite{harrison2013coding}. In addition, puncturing technique is an efficient way for hiding the secret part of the message \cite{5740591}. In the puncturing procedure, some bits of the codewords are eliminated at the transmitter.

\par Using punctured LDPC codes for providing security has been investigated in several studies \cite{5740591,6810282,5750045}. A random puncturing with concealed locations from Eve, satisfies the weak secrecy condition for the binary erasure wiretap channel \cite{6810282}. A random puncturing of secret messages also results in high equivocation rate at Eve by assigning some constraints for the Gaussian wiretap channel \cite{5750045}.

\par It should be noted that LDPC codes belong to the category of graph based codes, i.e., codes with a graphical representation, which have been tremendously utilized in the last two decades due to their superior performance in achieving the capacity of the most known channels. It has been shown that for an LDPC code with code length $n$, the complexity of its decoder is an order of $O(n)$ compared to that of Polar codes which is an order of $O(n\log{n})$ \cite{7264976}. Additionally, performance of LDPC codes under puncturing can be optimized for various applications \cite{bloch2011physical}, however to the best of our knowledge, no such optimal solution for the puncturing of polar codes is proposed so far.

\par Therefore, as mentioned earlier, in this paper we aim to propose a coding scheme using LDPC codes that can provide the physical layer security of GMAC-WT. We implement a nested structure by considering some random bits as random messages at the encoders. Therefore, the encoding process is stochastic from the secret messages standpoint. Furthermore, we conceal the secret messages by puncturing associated bits in the codewords. In this regard, there are two main concerns in puncturing procedure. First, a mother LDPC code for the puncturing should be selected among the optimized ensembles. In this case, low information leakage rate is obtained at Eve by utilizing a capacity approaching mother code \cite{harrison2013coding}. Second, the way of choosing the bits for puncturing affects the security gap and the minimum required SNR at the legitimate receiver. For this issue, we designate an optimization problem, which maximizes the secure rate and it determines the optimal puncturing distribution. Moreover, we perform a random puncturing to compare its performance with that of the optimized puncturing scheme. Note that, we assume that Eve has perfect knowledge about the index of punctured bits in the transmitted codewords.

\par We simulate the proposed secure encoder to illustrate the results of our scheme for the practical finite block length. We utilize the security gap as a security measure. As mentioned before, the security gap is obtained
from BER performance of the code, thus it is an appropriate metric over GMAC-WT. We are interested  in attaining half BER at Eve for admissible security gap which is the same as what has been assumed in  \cite{5740591} for the Gaussian wiretap channel.
\par According to our simulation results, the security gap performance has been improved using the punctured LDPC codes (compared to an unpunctured codes). As expected, the optimized puncturing scheme results in the least security gap. Moreover, our simulation results show the reduction of the security gap for random puncturing scheme. We study both equal and unequal power conditions at the transmitters. Consequently, unequal secure rates corresponding to the unequal transmit power of the transmitters are addressed. In all cases, significant improvement of the security gap is observed.

\par The rest of this paper is organized as follows. In Section \ref{sec-pre}, some preliminary concepts are provided. We propose a secure encoding method based on puncturing the confidential messages in Section \ref{sec-encoding}. Modified form of the joint decoder and EXIT chart for punctured LDPC codes are also described in this section. The last subsection of Section \ref{sec-encoding}, explains the optimization procedure of puncturing distribution on GMAC-WT. Simulation results of the proposed encoding scheme are illustrated in Section \ref{sec-simul}. Finally, the paper is concluded in Section \ref{sec-concul}. Throughout this paper, upper-case italic and upper-case bold-faced letters indicate
random variables and matrices, respectively. Additionally, lower-case bold-faced letters refer to
vectors. We also use $Y^n$ to indicate the vector $(Y_1,...,Y_n)$.

\section{{Preliminaries}}
\label{sec-pre}
\subsection{System model}
%%%%%%%%%%%%%%%%%%%%%%%%%%%%%%%%%%%%%%% Figure 1 %%%%%%%%%%%%%%%%%%%%%%%%%%%%%%%%%%%%%%%%%
\begin{figure}
\resizebox{\hsize}{!}{
\centering
\begin{tikzpicture}
[scale=0.8,node distance=15mm and 30mm,
terminal/.style={
% The shape:
rectangle,
% The size:
% The border:
rounded corners=2mm,
ultra thick,
draw=black!100,
% The filling:
top color=white, % a shading that is white at the top...
bottom color=gray!10!white!50, % and something else at the bottom
% Font
font=\itshape
}]
\node (encoder1) [terminal,minimum size=10mm,rounded corners=1mm] {  Encoder\textsubscript{1}  };
\node (encoder2) [terminal,minimum size=10mm,rounded corners=1mm,below = of encoder1] {  Encoder\textsubscript{2}  };
\node (plus1) [terminal,minimum size=2mm,rounded corners=2.5mm,right= of encoder1] {+};
\node (plus2) [terminal,minimum size=2mm,rounded corners=2.5mm,right= of encoder2] {+};
\node (decoder1) [terminal,minimum size=10mm,rounded corners=1mm,right= 15mm of plus1 ] {  Decoder\textsubscript{1}  };
\node (decoder2) [terminal,minimum size=10mm,rounded corners=1mm,,right= 15mm of plus2] {  Decoder\textsubscript{2}  };
\path  (encoder1) edge[line width=0.4mm,->] (plus1)
(plus1) edge[line width=0.4mm,->] ( decoder1) ($ (plus1.east) + (9mm,3mm)  $) node{\Large$\emph{Y}^n$} ;
\path  (encoder2) edge[line width=0.4mm,->] (plus2)
(plus2) edge[line width=0.4mm,->] ( decoder2) ($ (plus2.east) + (9mm,3mm)  $) node{\Large$\emph{Z}^n$} ;
\draw  [line width=0.4mm,->]  ($ (encoder1.east) + (8mm,4mm) $) node{$\emph{X}^{n,[1]}$}
($ (encoder1.east) + (16mm,0) $)
--
 ($ (plus2.west)  $);
\draw [line width=0.4mm,->]   ($ (encoder2.east) + (8mm,4mm) $) node{$\emph{X}^{n,[2]}$}
($ (encoder2.east) + (16mm,0) $)
--
 ($ (plus1.west)  $);
\draw [line width=0.4mm,->]
($ (encoder1.west) - (9mm,-4mm) $) node{\Large$\emph{M}^{[1]}$}
($ (encoder1.west) - (10mm,0) $)
--
 ($ (encoder1.west)  $);
\draw [line width=0.4mm,->]
($ (encoder2.west) - (9mm,-4mm) $) node{\Large$\emph{M}^{[2]}$}
($ (encoder2.west) - (10mm,0) $)
--
 ($ (encoder2.west)  $);
\draw [line width=0.4mm,->] ($ (decoder1.east) + (10mm,4mm)  $) node{\large$\emph{M}^{[1]}$\,$\emph{M}^{[2]}$}
($ (decoder1.east)  $)
--
 ($ (decoder1.east) + (20mm,0)  $);
\draw [line width=0.4mm,->] ($ (decoder2.east) + (10mm,4mm)  $) node{\large$\cancel{\emph{M}}^{[1]}$\,$\cancel{\emph{M}}^{[2]}$}
($ (decoder2.east)  $)
--
 ($ (decoder2.east) + (20mm,0)  $);
\draw [gray,thick,dashed,rounded corners=2mm](6.6cm,-0.5cm)-- (6.6cm,1.5cm) -- (11.1cm,1.5cm) -- (11.1cm,-1.4cm)-- (6.6cm,-1.4cm)--(6.6cm,-0.5cm)
(9cm,2cm) node [text=gray]{\Large \emph{Bob}};
\draw [gray,thick,dashed,rounded corners=2mm](6.6cm,-4cm)-- (6.6cm,-1.6cm) -- (11.1cm,-1.6cm) -- (11.1cm,-4.5cm)-- (6.6cm,-4.5cm)--(6.6cm,-4cm)
(9cm,-5cm) node [text=gray]{\Large \emph{Eve}};
\draw [line width=0.4mm,->]
($ (plus1.north) + (0,10mm) $) node{$\emph{N}_B^{n}$}
($ (plus1.north) + (0,7mm) $)
--
 ($ (plus1.north)  $);
\draw [line width=0.4mm,->]
($ (plus2.north) + (0,10mm) $)  node{$\emph{N}_E^{n}$}
($ (plus2.north) + (0,7mm) $)
--
 ($ (plus2.north)  $);
\end{tikzpicture}
}
\caption{A block diagram of a Gaussian multiple access wiretap channel (GMAC-WT).} \label{F1}
\end{figure}
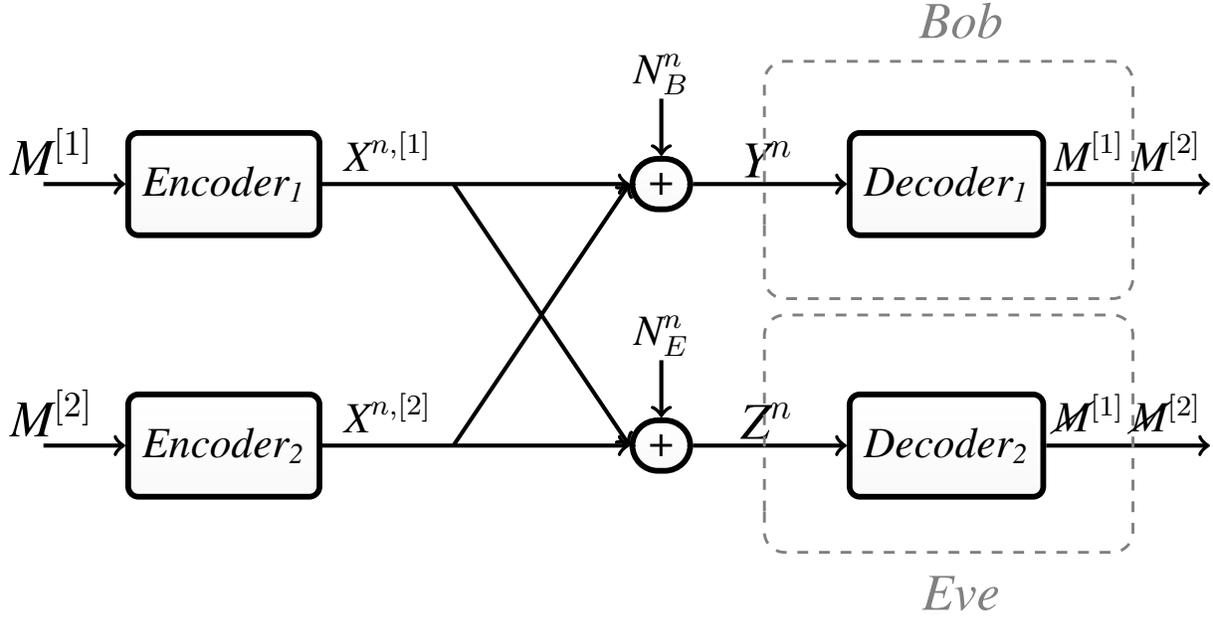
%%%%%%%%%%%%%%%%%%%%%%%%%%%%%%%%%%%%%%%% End Figure 1 %%%%%%%%%%%%%%%%%%%%%%%%%%%%%%%%%%%%%

\par The GMAC-WT system contains two users with confidential messages denoted by $\emph{M}^{[1]}$ and $\emph{M}^{[2]}$, as shown in Fig.~\ref{F1}. The BPSK modulated codewords denoted by $\emph{X}^{n,[1]}$ and $\emph{X}^{n,[2]}$ are the inputs of the channel. The relations between the inputs and the outputs of GMAC-WT are given by:
\begin{subequations}
\label{channel_eq}
\begin{align}
&\emph{Y}^n=\sqrt{p_1}\emph{X}^{n,[1]}+\sqrt{p_2}\emph{X}^{n,[2]}+\emph{N}_B^{n},\\
&\emph{Z}^n=\sqrt{p_1}\emph{X}^{n,[1]}+\sqrt{p_2}\emph{X}^{n,[2]}+\emph{N}_E^{n},
\end{align}
\end{subequations}
where $\emph{Y}^n$ and $\emph{Z}^n$ are the channel outputs at Bob and Eve, respectively. Moreover, $\emph{N}_B^{n}$ and $\emph{N}_E^{n}$ are individually independent and identically distributed (i.i.d) zero-mean Gaussian noises with $\sigma_B^2$ and $\sigma_E^2$ variances. The former corresponds to the main channel and the latter to the wiretap channel. However, these noises are statistically independent from each other and $\sigma_E^2$ is assumed to be greater than $\sigma_B^2$.
By assigning power constraints $p_1$ and $p_2$ corresponding to each user, two equal and unequal transmit power cases happen at the transmitters.

\subsection{LDPC codes}
\par LDPC codes are a class of linear block codes defined by a sparse parity-check matrix $\textbf{H}$ \cite{richardson2008modern}. It is represented by a bipartite graph called the Tanner graph consisting of variable nodes and check nodes corresponding to the codeword bits and the parity-check bits, respectively. Decoding of LDPC codes is performed by exchanging the probabilistic messages between the mentioned nodes. The belief propagation (BP) algorithm implements the decoding process with a good performance and a tolerable complexity \cite{richardson2008modern}.
\par Tanner graph can be described by two polynomials denoted by $\lambda(x)=\sum _{i=2} ^{D_{v}} \lambda_i x^{i-1}$ and $\rho(x)=\sum _{i=2} ^{D_c} \rho_i x^{i-1}$ which determine the degree distributions of variable and check nodes from an edge perspective, where $D_v$ and $D_c$ are maximum degrees of variable and check nodes, respectively. $\lambda_i$ and $\rho_i$ are used to denote fraction of edges connected to degree $i$ variable and check nodes, respectively. The code rate $R$ associated to the pair $\left(\lambda(x),\rho(x)\right)$ is defined as follows:
\begin{align}
\label{code_rate}
&R = 1 - \frac{{\sum_{i=2}^{D_c} {{{{\rho _i}} \mathord{\left/
 {\vphantom {{{\rho _i}} i}} \right.
 \kern-\nulldelimiterspace} i}} }}{{\sum_{i=2}^{D_v} {{{{\lambda _i}} \mathord{\left/
 {\vphantom {{{\lambda _i}} i}} \right.
 \kern-\nulldelimiterspace} i}} }}.
\end{align}
In addition, $L(x)=\sum _{i=2} ^{D_{v}} L_i x^{i}$ and $R(x)=\sum _{i=2} ^{D_{c}} R_i x^{i}$ are node perspective degree distributions, where $L_i$ and $R_i$  are the fraction of degree $i$ variable and check nodes, respectively.
\par In the puncturing technique that can be applied on LDPC codes, some of the encoded bits are not sent over the channel \cite{6365873}. Let $\pi(x)=\sum_{i=2}^{Dv} \pi_i x^{i-1}$, denote a distribution function of puncturing that $\pi_i$ implies the fraction of variable nodes of degree $i$ that should be punctured in a codebook. Therefore, puncturing rate is characterized as $R_p=\sum _{i=2} ^{D_{v}} L_i \pi_i$. For the random puncturing, $\pi_i$ with $2\le i \le D_v$ is equal to a constant $R_p$ over all degrees of variable nodes. It means that an equal fraction of all degrees of variable nodes are selected to be punctured, and hence,
\begin{align}
\label{rand_punc_distribution}
&\pi(x)=R_p\times\sum _{i=2} ^{D_v} x^{i-1}.
\end{align}
\par There are two major asymptotic analysis tools for LDPC codes known as density evolution (DE) and  extrinsic information transfer (EXIT) chart \cite{richardson2008modern}. EXIT chart analysis is based on the Gaussian approximation and it is an efficient tool for determining decoding thresholds and thus providing the optimum degree distribution of LDPC codes \cite{richardson2008modern}.

\section{Secure encoding based on punctured LDPC codes}\label{sec-encoding}

\par In this section, we first describe a nested structure for the secure encoders over GMAC-WT, and then explain the joint decoder on LDPC codes. Additionally, we propose a modified form of the joint decoder and EXIT chart analysis, corresponding to the case of punctured LDPC codes. The optimization problem defined in this section results in the optimal distribution of puncturing. The last subsection is on the security gap as a measure of secrecy.

\subsection{Construction of secure encoder}
\par
\begin{figure}
\centering
\resizebox{\hsize}{!}{
\begin{tikzpicture}
[node distance=15mm and 25mm,
terminal/.style={align=center,  text centered,
% The shape:
rectangle,
% The size:
% The border:
ultra thick,
draw=black!100,
% The filling:
top color=white, % a shading that is white at the top...
bottom color=gray!10!white!50, % and something else at the bottom
% Font
font=\itshape
}]
\node  (encoder) [terminal,minimum size=25mm,rounded corners=1mm,text width=25mm,] {\hspace{0.5mm}\Large LDPC Systematic Encoder};
\node (puncture) [terminal,minimum size=15mm,rounded corners=1mm,right = of encoder,text width=30mm] {\Large Message bits Puncturer};
\node (rand_gen) [terminal,minimum size=5mm,rounded corners=1mm ,below left = 1mm and 4.5mm of  encoder,text width=20mm] { Random Message};
\path  (encoder) edge[line width=0.6mm,->]   (puncture);
\draw  ($(encoder.east) + (12mm,4mm) $) node{ $[\textbf{p}^{[j]} \,\textbf{m}^{\prime[j]} \, \textbf{m}^{[j]}]$};
\draw  [line width=0.6mm,->]  ($ (rand_gen.north) + (-5mm,5mm) $) node{$\textbf{m}^{\prime[j]}$}
($(rand_gen.north)$)
--
 ($ (encoder.west) + (-16mm,0mm) $);
\draw [line width=0.6mm,->]   ($ (puncture.east) + (10mm,4mm) $) node{ $[\textbf{p}^{[j]} \,\textbf{m}^{\prime[j]}]$}
($ (puncture.east)$ )
--
 ($ (puncture.east) +(18mm,0)  $);
\draw [line width=0.6mm,->]   ($ (encoder.west) - (10mm,-4mm) $) node{ $[\textbf{m}^{\prime[j]} \, \textbf{m}^{[j]}]$}
($ (encoder.west) -(16mm,0)  $)
--
 ($ (encoder.west)   $);
\draw [line width=0.6mm,->]
($ (encoder.west) -(38mm,-2mm)  $) node{ $\textbf{m}^{[j]}$}
($ (encoder.west) -(38mm,-9mm)  $) node[text width=15mm,text centered,align=center]{\large \emph{ Secret\\Message}}
($ (encoder.west) -(35mm,0)  $)
--
 ($ (encoder.west) -(16mm,0)  $);
\draw [gray,thick,dashed,rounded corners=4mm](-4.2cm,-1cm)-- (-4.2cm,3cm) -- (7.3cm,3cm) -- (7.3cm,-3cm)-- (-4.2cm,-3cm)--(-4.2cm,-1cm)
(1.5cm,2.5cm) node [text=gray]{\Large \emph{Secure Encoder}};
\end{tikzpicture}
}
\caption{Secure encoding based on puncturing confidential messages of the user $j$.} \label{F2}
\end{figure}
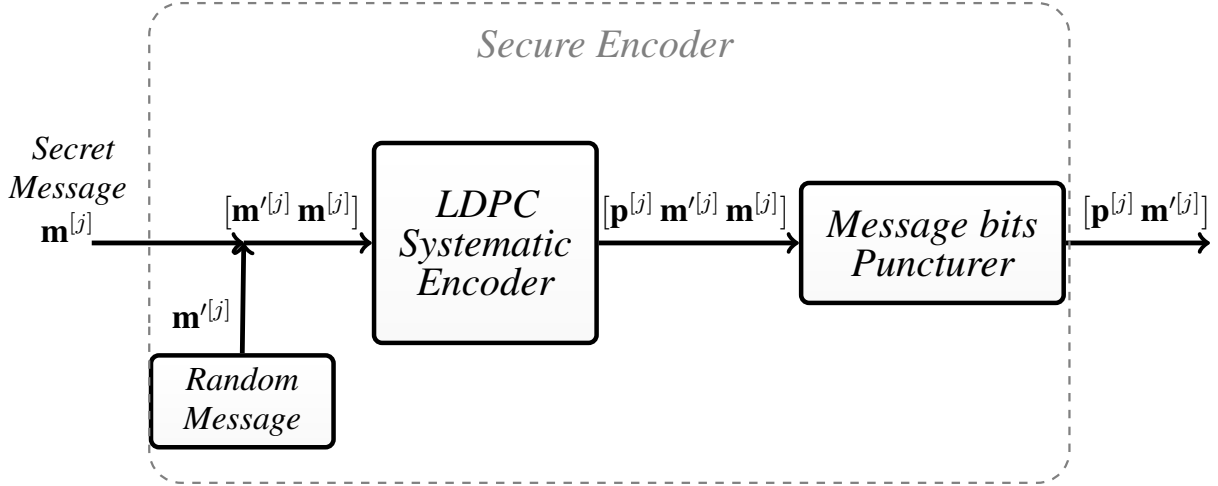

 Consider an LDPC code $\mathcal{C}^{\prime[j]}$, where $j=1,2$ implies $j$-th encoder, with a codeword length $n^\prime_j$, and a message length $l_j$. Let $k_j$ be the number of the message bits that should be confidential. Hence, a systematic codeword $\textbf{x}^{\prime[j]}$ of $\mathcal{C}^{\prime[j]}$ can be decomposed as $\textbf{x}^{\prime[j]}=[\textbf{p}^{[j]}\,\textbf{m}^{\prime[j]}\,\textbf{m}^{[j]}]$, where $\textbf{m}^{[j]}$, $\textbf{m}^{\prime[j]}$, and $\textbf{p}^{[j]}$ represent a $k_j$-bit confidential message, $(l_j-k_j)$-bit random message, and a $(n^{\prime}_j-l_j)$-bit parity message, respectively. Let also $\mathcal{C}^{[j]}$ be a codebook which is attained by puncturing $k_j$ bits of each codeword of $\mathcal{C}^{\prime[j]}$. Hence, the length of codewords of $\mathcal{C}^{[j]}$, denoted by $n_j$, equals $n^{\prime}_j-k_j$. The codebook $\mathcal{C}^{[j]}$ only contains $2^{k_j}$ number of punctured codewords of $\mathcal{C}^{\prime[j]}$. Therefore , punctured codewords of $\mathcal{C}^{\prime[j]}$, denoted by $[\textbf{p}^{[j]}\,\textbf{m}^{\prime[j]}]$, designate each confidential message $\textbf{m}^{[j]}$ in $\mathcal{C}^{[j]}$. Accordingly, $\mathcal{C}^{[j]}$ uses a nested construction with $\mathcal{C}^{\prime[j]}$, i.e., $\mathcal{C}^{[j]}\subset \mathcal{C}^{\prime[j]}$, and it consists a stochastic vector $\textbf{m}^{\prime[j]}$ which is comprised of random bits. Secure encoding procedure is organized as illustrated in Fig.~\ref{F2}. To encode a secret message $\textbf{m}^{[j]}$, $(l_j-k_j)$ bits should be randomly generated as $\textbf{m}^{\prime[j]}$, so as to $[\textbf{m}^{\prime[j]}\,\textbf{m}^{[j]}]$ become an input of the systematic LDPC encoder of $\mathcal{C}^{\prime[j]}$, as described in \cite{richardson2008modern}. This results in $\textbf{x}^{\prime[j]}=[\textbf{p}^{[j]}\,\textbf{m}^{\prime[j]}\,\textbf{m}^{[j]}]$ to be the output of the decoder. Finally, the encoded codeword $\textbf{x}^{[j]}=[\textbf{p}^{[j]}\,\textbf{m}^{\prime[j]}]$ will be transmitted by encoder $j$ after puncturing the message bits.

\emph{Mother code rate}, denoted by $R_{m}^{[j]}$, represents code rate of the mother code. \emph{Puncturing rate}, which is denoted by $R_p^{[j]}$, states the ratio of puncturing bits to the codeword length in a mother LDPC code. \emph{Secure rate}, denoted by $R_s^{[j]}$, is defined as the ratio of the number of secret message bits to the number of transmitted bits over the channel. Finally, $R_d^{[j]}$ denotes the \emph{design rate} which indicates the code rate of all transmitted message bits. In the above-mentioned parameters, the superscript $j$ denotes user $j$, $j=1,2$. The latter designing parameters are calculated as follows:
\begin{subequations}
	\label{rates}
\begin{align}\label{rate_R_m}
R_{m}^{[j]}&=\frac{l_j}{n^\prime_j},\\\label{rate_R_p}
R_p^{[j]}&=\frac{k_j}{n^\prime_j}=\frac{R_s^{[j]}}{1+R_s^{[j]}},\\\label{rate_R_s}
R_s^{[j]}&=\frac{k_j}{n_j}=\frac{R_p^{[j]}}{1-R_p^{[j]}},\\\label{rate_R_d}
R_d^{[j]}&=\frac{l_j}{n_j}=\frac{R_{m}^{[j]}}{1-R_p^{[j]}},
\end{align}
\end{subequations}
where $j=1,2$.

%%%%%%%%%%%%%%%%%%%%%%%%%%%%%%%%%%%%%%% Figure 2 %%%%%%%%%%%%%%%%%%%%%%%%%%%%%%%%%%%%%%%%%
%%%%%%%%%%%%%%%%%%%%%%%%%%%%%%%%%%%%%%%% End Figure 2 %%%%%%%%%%%%%%%%%%%%%%%%%%%%%%%%%%%%%
\subsection{LDPC joint decoder over GMAC}
\par An efficient joint decoder on multi-user systems was first implemented in \cite{1023606} using message passing algorithm. LDPC codes with BP decoder are appropriate for implementing by a joint decoder. This approach has been utilized over a two-user GMAC in \cite{roumy2007characterization} and \cite{balatsoukas2012design}. Their proposed joint decoding is described by a two-user Tanner graph which consists of two single-decoders corresponding to the transmitters. In addition to variable and check nodes, another type of nodes exists in the two-user Tanner graph, called as state nodes. State nodes send the information of the single-decoders to each other, in each iteration of the joint decoder. Therefore, the nodes of the joint decoder exchange two types of messages among themselves: intra-decoder messages exchanged among the variable and the check nodes of each single-decoder, and inter-decoder messages exchanged among the state and the variable nodes of both single-decoders. Intra-decoder messages are associated to the BP algorithm, so based on \cite{balatsoukas2012design}, they have the same update rules as represented in (\ref{BP_update_rule}) for $j=1,2$,
\begin{equation}\label{BP_update_rule}
\begin{aligned}
L_{{v_i} \to {c_k}}^{[j]} &= L_{{s_i} \to {v_i}}^{[j]} + \sum\limits_{c_{k'} \in {N({v_i})\backslash c_k}} {L_{{c_{k'}} \to {v_i}}^{[j]}} ,\\
L_{{c_k} \to {v_i}}^{[j]} &= \frac{1}{2}{\tanh ^{ - 1}}\left( {\prod\limits_{v_{i'} \in N({c_k})\backslash v_i} {\tanh \left( {\frac{{L_{{{v}_{i'}} \to {c_k}}^{[j]}}}{2}} \right)} } \right),
\end{aligned}
\end{equation}
where $L_{{v_i} \to {c_k}}^{[j]}$ ($L_{{c_k} \to {v_i}}^{[j]}$) denotes extrinsic information from $i$-th variable node ($k$-th check node) to $k$-th check node ($i$-th variable node) in the single-decoder corresponding to the $j$-th user. Note that, we use ${N(v_i)\backslash c_k}$ to determine the neighborhood of node $v_i$ in the Tanner graph, excluding $c_k$. Moreover, $L_{{s_i} \to {v_i}}^{[j]}$ is the inter-decoder message from $i$-th state node to $i$-th variable node which is a function of $L_{{v_i} \to {s_i}}^{[j']}$ and the received bit $y_i$, and is obtained by (\ref{state_node_rule1}) for the first user.
\begin{equation}\label{state_node_rule1}
\footnotesize
\begin{aligned}
&L_{{s_i} \to {v_i}}^{[1]} = f\left( {L_{{v_i} \to {s_i}}^{[2]},y_i} \right)  = \\
&\log \left( {{\textstyle{{\exp \left( { - {\textstyle{{{{\left( {y_i - \sqrt {{p_1}}  - \sqrt {{p_2}} } \right)}^2}} \over {2{\sigma ^2}}}} + L_{{v_i} \to {s_i}}^{[2]}} \right) + \exp \left( { - {\textstyle{{{{\left( {y_i - \sqrt {{p_1}}  + \sqrt {{p_2}} } \right)}^2}} \over {2{\sigma ^2}}}}} \right)} \over {\exp \left( { - {\textstyle{{{{\left( {y_i + \sqrt {{p_1}}  - \sqrt {{p_2}} } \right)}^2}} \over {2{\sigma ^2}}}} + L_{{v_i} \to {s_i}}^{[2]}} \right) + \exp \left( { - {\textstyle{{{{\left( {y_i + \sqrt {{p_1}}  + \sqrt {{p_2}} } \right)}^2}} \over {2{\sigma ^2}}}}} \right)}}}} \right),\\
&L_{{v_i} \to {s_i}}^{[1]} = \sum\limits_{c_{k'} \in N({v_i})} {L_{{c_{k'}} \to {v_i}}^{[1]}}.
\end{aligned}
\end{equation}
\par State node updating rule for the second user is established in the same way as follows:
\begin{equation}\label{state_node_rule2}
\footnotesize
\begin{aligned}
&L_{{s_i} \to {v_i}}^{[2]} = f\left( {L_{{v_i} \to {s_i}}^{[1]},y_i} \right)  = \\
&\log \left( {{\textstyle{{\exp \left( { - {\textstyle{{{{\left( {y_i - \sqrt {{p_2}}  - \sqrt {{p_1}} } \right)}^2}} \over {2{\sigma ^2}}}} + L_{{v_i} \to {s_i}}^{[1]}} \right) + \exp \left( { - {\textstyle{{{{\left( {y_i - \sqrt {{p_2}}  + \sqrt {{p_1}} } \right)}^2}} \over {2{\sigma ^2}}}}} \right)} \over {\exp \left( { - {\textstyle{{{{\left( {y_i + \sqrt {{p_2}}  - \sqrt {{p_1}} } \right)}^2}} \over {2{\sigma ^2}}}} + L_{{v_i} \to {s_i}}^{[1]}} \right) + \exp \left( { - {\textstyle{{{{\left( {y_i + \sqrt {{p_2}}  + \sqrt {{p_1}} } \right)}^2}} \over {2{\sigma ^2}}}}} \right)}}}} \right),\\
&L_{{v_i} \to {s_i}}^{[2]} = \sum\limits_{c_{k'} \in N({v_i})} {L_{{c_{k'}} \to {v_i}}^{[2]}}.
\end{aligned}
\end{equation}
\par The joint decoding procedure may be performed in a parallel or a serial schedule \cite{roumy2007characterization}. In this paper the parallel schedule is considered.
\subsection{Joint decoder for punctured LDPC code}
\par Because of puncturing some bits of the codeword, we propose a modified form of the joint decoder to reconstruct them. Punctured bits in each encoder have no interference on the transmitted sequence of the other one. Therefore, we attribute state nodes to the unpunctured bits that are transmitted over the channel. For the case that two users have equal power, there is no difference between the users, so the rates, code ensembles and puncturing distributions must be the same at both encoders. In this case, the number of unpunctured bits are equal for both users. Therefore, the forward messages of the $l$-th state node obtained from (\ref{state_node_rule1}) (or (\ref{state_node_rule2})) are associated to the $l$-th unpunctured variable node of user 1 (or user 2). Thus, the variable node update rule for unpunctured and punctured bits are as follows, respectively:
\begin{subequations}\label{punc_unpunc_VN_rule}
\begin{align}
\label{unpunc_VN_rule}
{L_{{v_i} \to {c_k}}^{[j]}}\Bigg|_{(1-\pi)}&=L_{{s_l} \to {v_i}}^{[j]} + \sum\limits_{c_{k'} \in N({v_i})\backslash c_k} {L_{{c_{k'}} \to {v_i}}^{[j]}},\\
\label{punc_VN_rule}
{L_{{v_i} \to {c_k}}^{[j]}}\Bigg|_{(\pi)} &=\sum\limits_{c_{k'} \in N({v_i})\backslash c_k} {L_{{c_{k'}} \to {v_i}}^{[j]}},
\end{align}
\end{subequations}
where subscript $(1-\pi)$ corresponds to the unpunctured bits and $(\pi)$ corresponds to punctured bits.
\par Different power constraints at the transmitters cause unequal rates and code ensembles for the encoders. Because of various puncturing rates, the codeword lengths are not the same at the output of the secure encoders. Therefore, different puncturing rates lead to an interference between more than one codeword of each user. To circumvent this problem, we assign different code lengths in LDPC systematic encoders such that the codewords have similar lengths in the output of the secure encoders. According to the $R_p^{[j]}$, $j=1,2$ and transmitted codeword lengths of both users ($n_1=n_2=n$), the mother code length $n_j^{\prime}$ is computed as follows:
\begin{align}
n_j^{\prime}=\frac{n}{\left(1-R_p^{[j]}\right)},\quad j=1,2.\nonumber
\end{align}
Finally, the decoding process is performed by considering message exchange of the unpunctured variable nodes with the state nodes using (\ref{punc_unpunc_VN_rule}).
\subsection{EXIT chart analysis on punctured LDPC codes }
\par Tracking the mutual information (MI) between extrinsic (or a-priori) messages of a decoder and its transmitted message is the principal part of the EXIT chart analysis. For this purpose, we assume that variable and check node messages have a consistent Gaussian distribution with mean $\sigma^2/2$ and variance $\sigma^2$ denoted by ${\mathcal{N}}\left( \sigma ^2/2,\sigma ^2 \right)$. Based on \cite{1291808}, the relation between MI and the variances of the exchanged messages is attained using $I=J(\sigma)$ and $\sigma=J^{-1}(I)$, where $J(.)$ function is given by:
\begin{equation}\label{J_functions}
\begin{aligned}
J\left( \sigma  \right) = 1 - {\textstyle{1 \over {\sqrt {2\pi } \sigma }}}&\int\limits_{\mathbb{R}} {\exp \left( { - {\textstyle{{{{\left( {l - {\sigma ^2}/2} \right)}^2}} \over {2{\sigma ^2}}}}} \right)}
{\log _2}\left( {1 + {e^{ { - l} }}} \right)dl.\nonumber
\end{aligned}
\end{equation}
Therefore, according to the (\ref{unpunc_VN_rule}), we obtain the variance of the extrinsic message for unpunctured  degree $i$ variable node, denoted by $ {\sigma _{Ev_i}^2}\Big|_{(1-\pi)}$, in (\ref{EXIT_var_unpunc}).
The extrinsic MI for unpunctured  degree $i$ variable node, denoted by $ {I_{Ev_i}}|_{(1-\pi)}$, is determined as (\ref{EXIT_MI_unpunc}).
\begin{subequations}
\begin{align}
\label{EXIT_var_unpunc}
&{\sigma _{Ev_i}^2} \Bigg|_{(1-\pi)}=\sigma _{Es_l}^2 + \left( {i - 1} \right)\sigma _{Av}^2,\\
\label{EXIT_MI_unpunc}
& {I_{Ev_i}}\Bigg|_{(1-\pi)}= J\left( {\sqrt {{(J^{ - 1}{{\left( {{I_{Es_l}}} \right)}})^2} + \left( {i - 1} \right){(J^{ - 1}{{\left( {{I_{Av}}} \right)}})^2}} } \right),
\end{align}
\end{subequations}
where $\sigma _{Es_l}^2$ ($I_{Es_l}$) is the variance (MI) associated to the extrinsic message of $l$-th state node and $\sigma _{Av}^2$ ($I_{Av}$) is the variance (MI) associated to a-priori message of the variable nodes.
\par As the updating rule of punctured variable nodes shows in (\ref{punc_VN_rule}), it only consists of messages from the check nodes. Thus, the variance and MI of the extrinsic message associated to the degree $i$ punctured variable node, denoted respectively by ${\sigma _{Ev_i}^2}\Big|_{(\pi)}$ and ${I_{Ev_i}} \Big|_{(\pi)}$, are the functions of the variance and MI of the a-priori message associated to the variable nodes, denoted respectively by $\sigma _{Av}^2$ and ${I_{Av}}$:
\begin{subequations}
\begin{align}
\label{EXIT_var_punc}
& {\sigma _{Ev_i}^2} \Bigg|_{(\pi)} =\left( {i - 1} \right)\sigma _{Av}^2,\\
\label{EXIT_MI_punc}
& {I_{Ev_i}} \Bigg|_{(\pi)} = J\left( {\sqrt { \left( {i - 1} \right){(J^{ - 1}{\left( I_{Av} \right)})^2}} } \right).
\end{align}
\end{subequations}
Therefore, we derive total EXIT function for the variable nodes as follows:
\begin{align}\label{EXIT_function}
I_{Ev}=\sum\limits_i {(\lambda _i(1-\pi_i) {I_{Ev_i}} \Bigg|_{(1-\pi)}+\lambda _i(\pi_i){I_{Ev_i}}\Bigg|_{(\pi)})}.
\end{align}
\par Based on (\ref{state_node_rule1}) and (\ref{state_node_rule2}), $I_{Es}$ is determined for each user as a function of the a-priori variance of the other user, denoted by $\sigma _{As}^{2}$, and the channel noise variance denoted by $\sigma_N^2$. Let $I_{Es}^{[1]}$ be for the first user as mentioned in (\ref{EXIT_var_punc_SVN}). For the second user, it is simply obtained by converting the user's index 1 to 2 and vice versa. Hence, it is omitted here for brevity.
\begin{equation}\label{EXIT_var_punc_SVN}
\footnotesize
\begin{aligned}
I_{Es}^{[1]} &= {\textstyle{1 \over 2}}J\left( {\sqrt {2F_ + ^{[1]}\left( {{\textstyle{1 \over 2}}\sum\limits_i {L_i^{[2]}i(1-\pi_i^{[2]}){(J^{ - 1}{\left( {I_{Av}^{[2]}} \right)})^2}} ,\sigma _N^2} \right)} } \right) \\
&+ {\textstyle{1 \over 2}}J\left( {\sqrt {2F_ - ^{[1]}\left( {{\textstyle{1 \over 2}}\sum\limits_i {L_i^{[2]}i(1-\pi_i^{[2]}){(J^{ - 1}{\left( {I_{Av}^{[2]}} \right)})^2}} ,\sigma _N^2} \right)} } \right).
\end{aligned}
\end{equation}
\par Note that, $F_ + ^{[j]}$ and $F_ - ^{[j]}$ are defined in (\ref{F_functions}) for user $j$, cf. at the bottom of the next page. $F_ + ^{[j]}$ is associated to the $+1$ and $+1$ messages and $F_ - ^{[j]}$ is associated to the $+1$ and $-1$ messages at the first and the second transmitters, respectively and are denoted in \cite{balatsoukas2012design}. Moreover, the superscripts $[j]$, in (\ref{EXIT_var_punc_SVN}), where $j=1,2$, for $L_i$, $\pi_i$ and $I_{Av}$ denote those for user $j$.
\begin{figure*}[b]\hrulefill
\begin{equation}\label{F_functions}
\begin{aligned}
F_ + ^{[1]}\left( \mu  \right) = &{\textstyle{1 \over {\sqrt \pi  }}}\int_{ - \infty }^{ + \infty } {e^{ - {z^2}} \log \left( {{\textstyle{{1 + \exp \left( {\sqrt {4\mu  + 8{p_2}} z + \mu  + 2{p_2}} \right)} \over {1 + \exp \left( { - \sqrt {4\mu  + 8{p_2}} z - \mu  - 2{p_2} - 4\sqrt {{p_1}} \sqrt {{p_2}} } \right)}}}} \right)} dz - \mu  + 2\left( {{p_1} - {p_2}} \right)\\
F_ - ^{[1]}\left( \mu  \right) = &{\textstyle{1 \over {\sqrt \pi  }}}\int_{ - \infty }^{ + \infty } {e^{ - {z^2}} \log \left( {{\textstyle{{1 + \exp \left( { - \sqrt {4\mu  + 8{p_2}} z - \mu  - 2{p_2}} \right)} \over {1 + \exp \left( {\sqrt {4\mu  + 8{p_2}} z + \mu  + 2{p_2} - 4\sqrt {{p_1}} \sqrt {{p_2}} } \right)}}}} \right)} dz + \mu  + 2{\left( {{{\sqrt p_1}} - \sqrt {{p_2}} } \right)^2}\\
F_ + ^{[2]}\left( \mu  \right) = &{\textstyle{1 \over {\sqrt \pi  }}}\int_{ - \infty }^{ + \infty } {e^{ - {z^2}} \log \left( {{\textstyle{{1 + \exp \left( {\sqrt {4\mu  + 8{p_1}} z + \mu  + 2{p_1}} \right)} \over {1 + \exp \left( { - \sqrt {4\mu  + 8{p_1}} z - \mu  - 2{p_1} - 4\sqrt {{p_1}} \sqrt {{p_2}} } \right)}}}} \right)} dz - \mu  + 2\left( {{p_2} - {p_1}} \right)\\
F_ - ^{[2]}\left( \mu  \right) = &{\textstyle{1 \over {\sqrt \pi  }}}\int_{ - \infty }^{ + \infty } {e^{ - {z^2}} \log \left( {{\textstyle{{1 + \exp \left( {\sqrt {4\mu  + 8{p_1}} z + \mu  + 2{p_1} - 4\sqrt {{p_1}} \sqrt {{p_2}} } \right)} \over {1 + \exp \left( { - \sqrt {4\mu  + 8{p_1}} z - \mu  - 2{p_1}} \right)}}}} \right)} dz - \mu  - 2{\left( {{{\sqrt p_2}} - \sqrt {{p_1}} } \right)^2}
\end{aligned}
\end{equation}
\end{figure*}
\par EXIT function of the check nodes is also the same as the single-decoder case \cite{1291808}, because of employing the same updating rules. Therefore, the extrinsic MI of check nodes, denoted by ${I_{Ec}}$, is as follows:
\begin{equation}
\begin{aligned}\label{EXIT_check}
{I_{Ec}} &= \sum\limits_k {{\rho _k}\left[ {1 - J\left( {\sqrt {\left( {k - 1} \right){(J^{ - 1}{\left( {1 - {I_{Ac}}} \right)})^2}} } \right)} \right]},
\end{aligned}
\end{equation}
where $I_{Ac}$ is the a-priori MI of the check nodes.
Since the extrinsic message of the check nodes is an a-priori message for the variable nodes, i.e., $I_{Ec}=I_{Av}$, a sufficient condition for successful decoding is $I_{Ac}<I_{Ev}$.

\subsection{Optimizing puncturing distribution}
\par It has been shown in \cite{6365873} and \cite{4069155} that puncturing the optimized degree distribution of LDPC codes has nearly equal performance as the mother code with the same code rate. Hence, we employ the optimized degree distribution of LDPC codes over GMAC as a mother code which performs close to the theoretical limits \cite{balatsoukas2012design}. Then, we obtain the optimized puncturing distribution for the considered mother code.
\par The optimization criterion is to maximize the puncturing rate which leads to the maximum secure rate, for each user. For a given decoding threshold, we can define a linear optimization problem to determine $\pi_i^{[j]}$ for $j=1,2$, as follows:
\begin{align}
\label{punc_opt_problem}
\max \limits_{\pi _i^{[j]}} \quad &\sum\limits_{i = 2}^{{D_v}} {L_i^{[j]}\pi_i^{[j]}} \\
\label{constraint1_punc}
\text{subject\,\,to}\quad&I_{Ac}^{[j]} < \sum\limits_{i = 2}^{{D_v}} {\lambda _i^{[j]}\pi _i^{[j]}{{{I_{Ev}^{i,[j]}} }\Bigg|_{(\pi)} }}\\ \nonumber
&\,\,+ \lambda _i^{[j]}\left( {1 - \pi _i^{[j]}} \right){{ {I_{Ev}^{i,[j]}} }\Bigg|_{\left( {1 - \pi } \right)}},\\
\label{constraint2_punc}
&\sum\limits_{i = 2}^{{D_v}} {L_i^{[j]}\pi_i^{[j]}}\leq R_m^{[j]},\\
\label{constraint3_punc}
&\pi _i^{[j]} \geq 0,i = 2,...,{D_v},\\
\label{constraint4_punc}
&\pi _i^{[j]} \leq 1,i = 2,...,{D_v},\\
\label{constraint5_punc}
&\pi _i^{[j]}=0,\left\{i|\lambda_i=0\right\}.
\end{align}
\par In the above, (\ref{constraint1_punc}) is the condition which leads to the decoder's convergence under the stable fixed-point $I_{Ev} = 1$ that corresponds to the zero error rate constraint. As mentioned before, the maximum amount of puncturing rate is equal to $R_m$, so the constraint (\ref{constraint2_punc}) is inevitable. The $\pi_i$ is the fraction of punctured degree $i$ variable nodes, thus it satisfies (\ref{constraint3_punc}), (\ref{constraint4_punc}) and (\ref{constraint5_punc}).
\par The optimization procedure is accomplished for each user, but the puncturing distribution of one user affects the optimization of the other user. Therefore, we start with an initial distribution for one user and perform the optimization procedure for the other one. Then, the obtained distribution is again used to optimize the next user. This process is repeated until the same result is achieved in the consecutive iterations.

\subsection{Security measurement metric}
\par To measure the level of secrecy in the proposed scheme, we use the security gap. As formerly mentioned, the security gap is defined based on BER of the secret message after decoding. Let $P_e^B$ and $P_e^E$ denote BER values of a secret message at Bob's and Eve's decoders, respectively. If Eve decodes the secret message with half error probability, she will not be able to extract any information about it. Hence, the variance of Eve's channel noise $\sigma_E^2$ should be greater than or equal to a minimum value $\sigma_{E_\text{min}}^2$. In spite of that, a reasonable level of reliability based on BER, should be established at the desired decoder. Therefore, the variance of Bob's channel noise $\sigma_B^2$ should be less than or equal to a maximum value $\sigma_{B_\text{max}}^2$. In summary, the following conditions on the reliability and the security should be satisfied to have a reliable and a secure communication.
\begin{subequations}
\label{error_pro_conditions}
\begin{align}
P_e^E \ge 0.5\quad \quad & \Rightarrow  \sigma_E^2 \ge \sigma_{E_\text{min}}^2,\\
P_e^B \le {10^{ - 5}}\quad  & \Rightarrow \sigma_B^2 \le \sigma_{B_\text{max}}^2.
\end{align}
\end{subequations}
\par The security gap denoted by ${g}$ is the difference of two limiting values of variances $\sigma_{E_\text{min}}^2$ and $\sigma_{B_\text{max}}^2$ in dB scale, and hence is given by:
\begin{align}
\label{Gap}
{g}=10 \log_{10}(\frac{\sigma_{E_\text{min}}^2}{\sigma_{B_\text{max}}^2}) \, \text{dB}. \nonumber
\end{align}
Conceptually, the security gap determines the quantity of an additional noise on the wiretap channel which is required to have the secure and the reliable communication, simultaneously. In this paper, we choose the security rate and the design rate of codes associated to the power constraints of the transmitters, such that the reliability (security) condition is satisfied for the same $\sigma_{B_\text{max}}^2$ ($\sigma_{E_\text{min}}^2$) value on the confidential (eavesdropped) message. Therefore, the security gap can be calculated based on SNR values as calculated in \cite{5740591}.

\section{Simulation results and discussion}
\label{sec-simul}
\par In order to demonstrate the performance of our proposed scheme, we simulate the proposed secure coding scheme in this section. We consider both equal and unequal transmit power cases.
\par For a given $R_s$ of each user, the desired $R_p$ is obtained from (\ref{rate_R_p}). We calculate the puncturing distribution in two ways:
\begin{itemize}
\addtolength{\itemindent}{4mm}
    \item By applying the optimal puncturing distribution derived by (\ref{punc_opt_problem}) for a proper decoding threshold which results in the desired $R_p$.
    \item By using a random puncturing distribution associated to $R_p$ derived by (\ref{rand_punc_distribution}).
\end{itemize}

\par It should be noted that the mother code rate $R_m$ determines length of the random message and the parity bits. Actually, as (\ref{rate_R_p}) shows, we can assign $R_m$ to its maximum value $R_p$ by eliminating the random message. Hence, in this case, the absence of random parameters in the encoder results in the maximum security gap. Moreover, selecting $R_m$ equals $R_p$ maximizes $R_d$ which leads to minimize the SNR loss.
On the contrary, increasing the number of random bits reduces the security gap. This also leads the parity bits to decrease, and accordingly $R_d$ which imposes an extra SNR loss on the performance of the code. If $R_m$ is equal to $R_s$, the number of random bits is sufficient to attain a low security gap, while the SNR loss is also admissible, cf. simulations in \cite{5740591}. Therefore, we also consider the equality of $R_m$ and $R_s$ in our simulations.

\par After selecting $R_m$, the optimized finite-length LDPC code, based on \cite{balatsoukas2012design}, is applied as a mother code. Then, the punctured codewords derived by an optimized or a random puncturing are transmitted over GMAC-WT. Bob and Eve extract secret messages from the received sequences at their receivers. Eventually, the required security gap is computed based on the amount of intended BERs at the receivers.
\par In the following simulations, we set the code lengths $n_1$ and $n_2$ equal $10^4$, and $P_{e,max}^B=10^{-5}$ in all examples.
\subsection{Equal power case}
\par We first impose an equal transmit power constraint corresponding to equal secure rates for both users.  The same mother codes are also applied for both users.
\par Let $p_1$ and $p_2$ equal $1$ and $R_{s}$ be $0.3333$. The optimized degree distributions of mother codes are as follows:
\begin{equation}
\label{equal_power_Dist}
\begin{aligned}
\lambda^{[1],[2]}(x)=&0.1993x +0.2796x^2+0.0096x^8+0.1814x^{10}\\
&\,+0.0113x^{15}+0.3188x^{99},\\
\rho^{[1],[2]}(x)=&x^6.
\end{aligned}
\end{equation}
\begin{figure}
\resizebox{\hsize}{!}{\includegraphics[scale=.3]{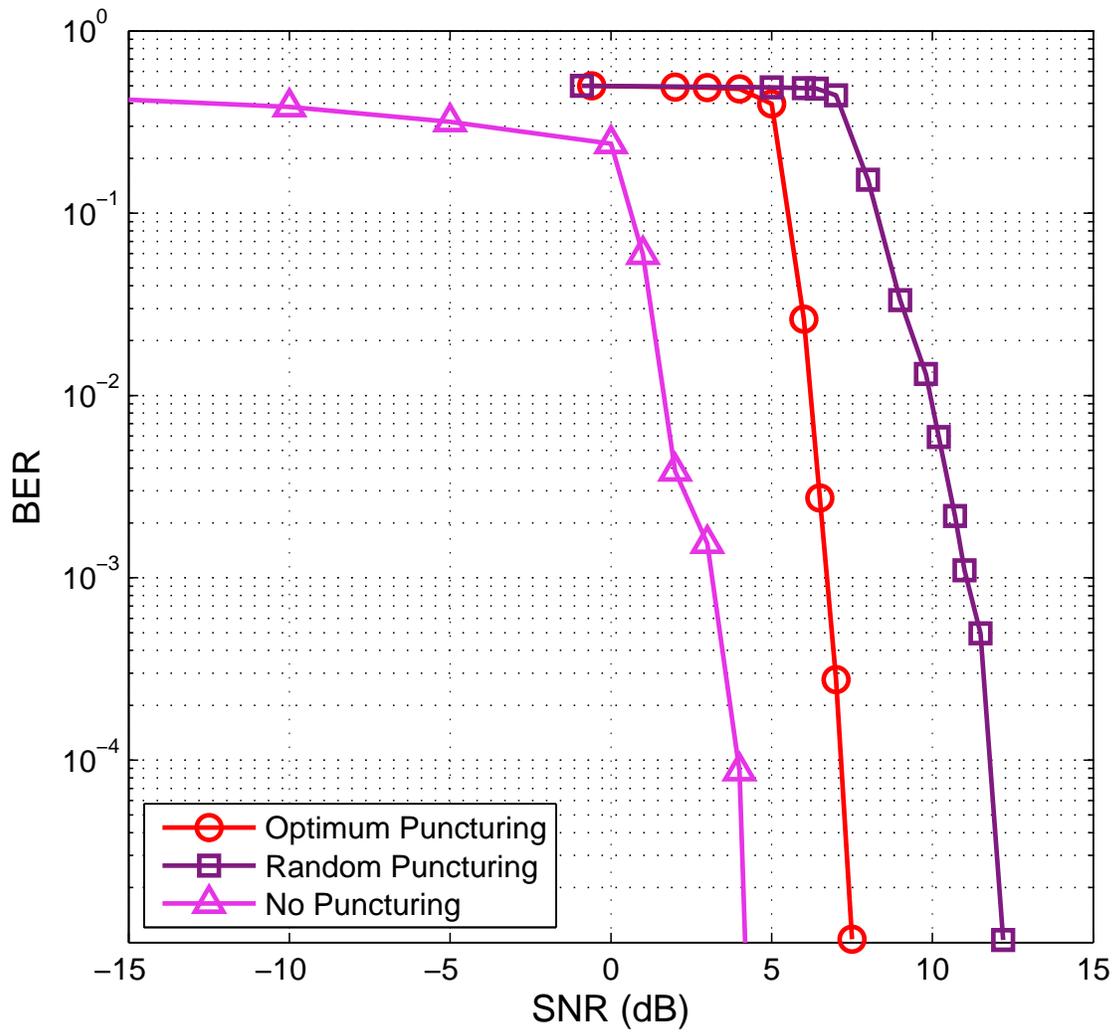}} \caption{BER versus SNR of equal power case, with the security rates $R_s^{[1]}=R_s^{[2]}=0.3333$ for the optimized punctured, the randomly punctured and unpunctured codewords.} \label{F_equal_BER}
\end{figure}
\begin{figure}[t]
\resizebox{\hsize}{!}{\includegraphics[scale=.3]{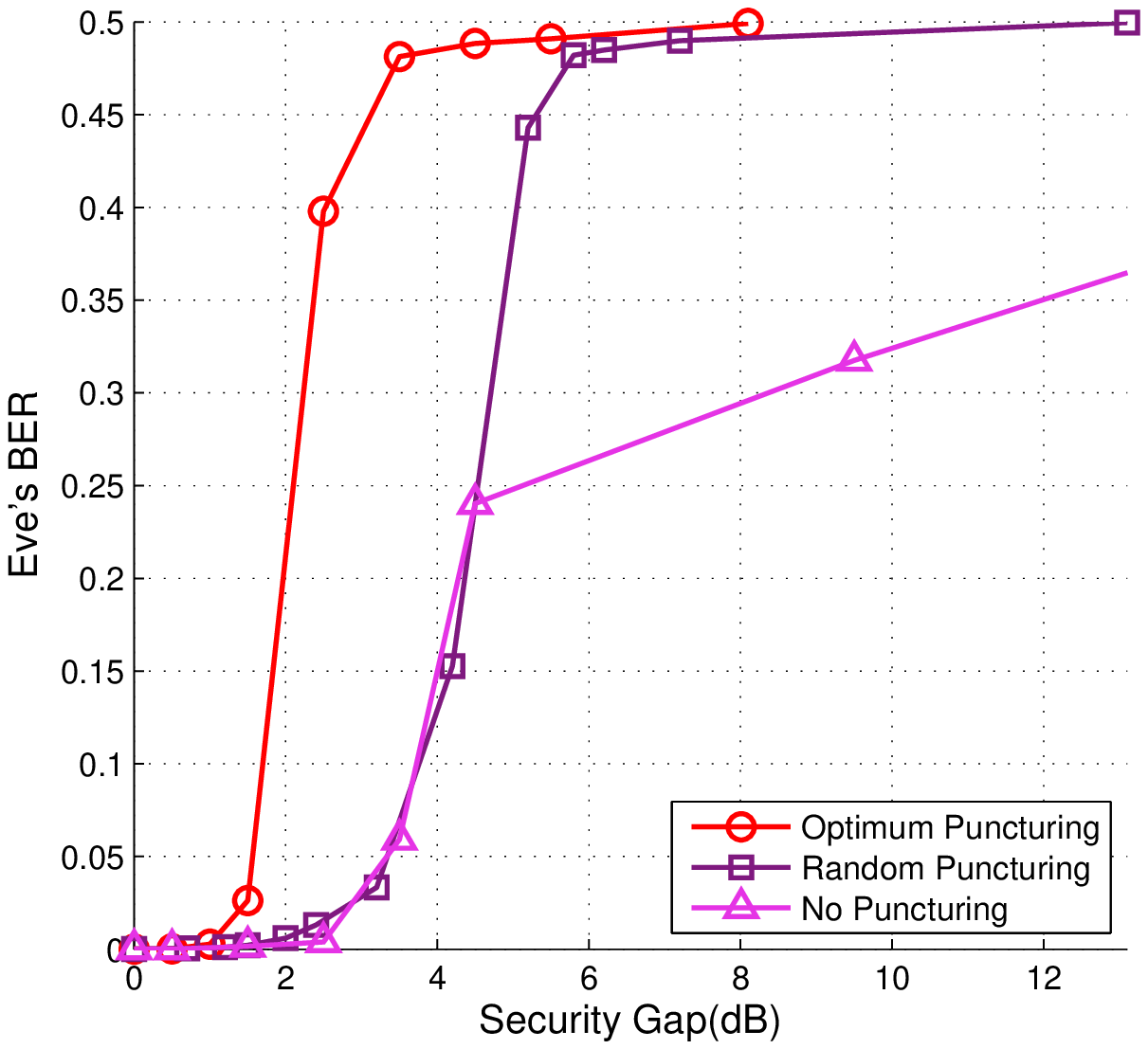}} \caption{Eve's BER versus the security gap of equal power case, with the security rates  $R_s^{[1]}=R_s^{[2]}=0.3333$, for the optimized punctured, randomly punctured and unpunctured codewords.} \label{F_equal_gap}
\end{figure}
\par According to $R_s$ and using (\ref{rate_R_p}), the desired puncturing rate $R_p$ is $0.25$. The desired $R_p$ is obtained by considering a decoding threshold equals $0.9151$ in the optimization procedure (\ref{punc_opt_problem}). Thus, the associated puncturing distributions for both encoders are given by:
\begin{equation}
\label{equal_power_punc_Dist}
\begin{aligned}
\pi^{[1],[2]}(x)=&0.283x +0.2723x^2.
\end{aligned}
\end{equation}
\par By applying (\ref{rate_R_d}) and for the above mentioned parameters, the design rate $R_d$ is equal to $0.4444$. We also apply a random puncturing distribution defined in (\ref{rand_punc_distribution}) for $R_p=0.25$ to compare its performance with that of the optimized puncturing distribution. In Fig.~\ref{F_equal_BER}, the BER of a standard encoder (without puncturing) is demonstrated in terms of SNR. Since there is the same BER performance for both users' messages, we only illustrate the performance of one code for the sake of simplicity.
\par Fig.~\ref{F_equal_gap} demonstrates the BER curves of Eve against the security gap. It confirms that our proposed secure encoder reduces the security gap significantly. Specially, using the optimized puncturing distributions at the secure encoders, increases the BER of Eve to get its highest value. If Eve's BER values $P_e^{E}$ are considered as $0.45$, $0.48$ and $0.49$, the security gap values will be $3$, $3.2$ and $5.5$ dB for the optimized punctured case, and $5.3$, $5.6$ and $7.2$ dB for the random punctured case, respectively. Moreover, the corresponding security gaps will be $22.3$, $30.3$ and $36.3$ dB for the unpunctured message case and the same BER values, respectively. This improvement is at the expense of increasing the minimum amount of Bob's SNR for attaining a reliable communication in the punctured case compared to the unpunctured one. This SNR difference, which is also called the SNR loss, means that the users should spend more power at the transmitters. As seen in Fig.~\ref{F_equal_BER}, the amount of the SNR loss is about $3.2$ dB for the optimized punctured case and $7.9$ dB for the random punctured case, respectively.

\par
\begin{figure}[t]
\resizebox{\hsize}{!}{\includegraphics[scale=.3]{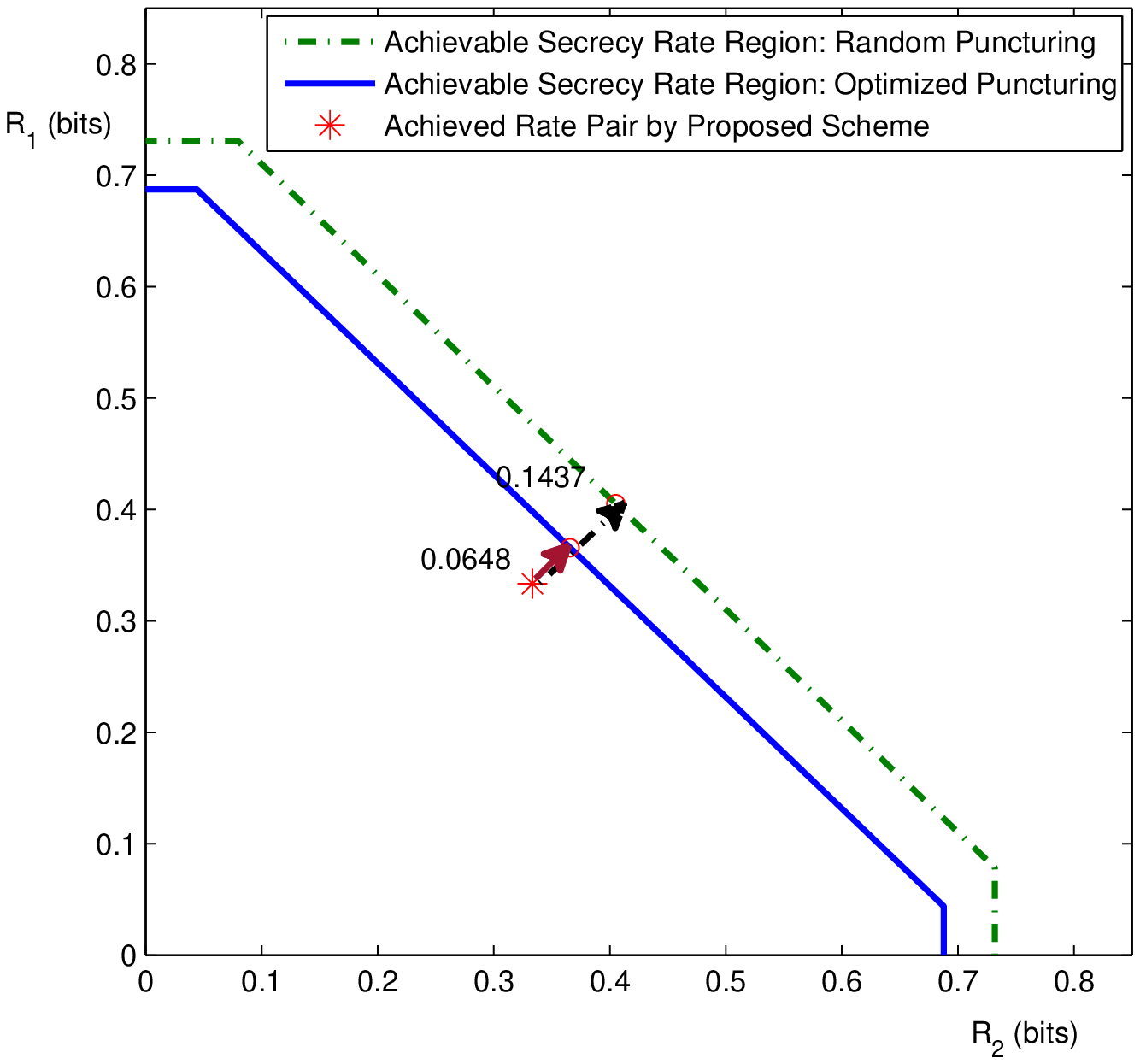}} \caption{Comparison between an achievable secrecy rate region and the obtained rate pair by applying the random and the optimized puncturing schemes for equal transmit powers.} \label{compared_rate_region_eq}
\end{figure}

 The best theoretical achievable secrecy rate region derived so far for the two user GMAC-WT was illustrated in \cite{EkremUlukus,iet2015.0468}. As it can be seen in Fig.~\ref{F_equal_BER}, $\sigma_B^2=0.1778$ and $\sigma_E^2=1.1482$ are calculated for the optimized puncturing case with $P_{e}^B=10^{-5}$, $P_e^E\simeq0.5$, $p_1=p_2=1$. Similarly, $\sigma_B^2=0.0603$ and $\sigma_E^2=1.2303$ are obtained for the random puncturing case. Thus, the achievable secrecy rate regions for the optimized and the random puncturing parameters are illustrated in Fig.~\ref{compared_rate_region_eq}. We also show the achieved rate pair using the proposed secure encoder in the same figure. The achieved sum-rate is $0.0648$ bits away from the maximum sum-rate on the secrecy rate region for the optimized puncturing case. Likewise, the difference between the maximum sum-rate on the secrecy rate region and the achieved sum-rate is about $0.1437$ bits for the random puncturing.

\subsection{Unequal power case}
\par Unequal transmit powers are considered in this part, where $p_1$ and $p_2$ are $1.5$ and $0.5$, respectively. Secure rates are also assumed as $R_s^{[1]}=0.4451$ and $R_s^{[2]}=0.2215$. Let $R_m=R_s$, then the optimized degree distributions for the mother code rates $R_m^{[1]}=0.4451$ and $R_m^{[2]}=0.2215$ associated to user 1 and user 2 are as follows, respectively:
\begin{equation}
\label{unequal_power_Dist1_[1]}
\begin{aligned}
\lambda^{[1]}(x)=&0.1559x +0.2974x^2+0.0394x^7+0.1305x^{8}\\
&\,+0.3768x^{99},\\
\rho^{[1]}(x)=&x^8,
\end{aligned}
\end{equation}
\begin{equation}
\label{unequal_power_Dist1_[2]}
\begin{aligned}
\lambda^{[2]}(x)=&0.1657x +0.2298x^2+0.0907x^6+0.0521x^{7}\\
&\,+0.4617x^{99},\\
\rho^{[2]}(x)=&x^6.
\end{aligned}
\end{equation}
\par For the given secure rates $R_s^{[1]}=0.4451$ and $R_s^{[2]}=0.2215$, the desired puncturing rates are calculated as $R_p^{[1]}=0.308$ and $R_p^{[2]}=0.1814$ using (\ref{rate_R_p}), respectively. The desired puncturing distributions are attained from the optimization procedure (\ref{punc_opt_problem}) by assigning the decoding threshold equals $0.8998$. The puncturing distributions for user 1 and user 2 are given by:
\begin{equation}
\begin{aligned}
\label{unequal_power_punc_Dist_[1]}
\pi^{[1]}(x)=&0.3431x +0.3029x^2+0.2391x^8+0.3865x^{99},\\
\pi^{[2]}(x)=&0.2828x +0.1239x^2+0.0774x^{99}.\\
\end{aligned}
\end{equation}
\par
\begin{figure}[t]
\resizebox{\hsize}{!}{\includegraphics[scale=.3]{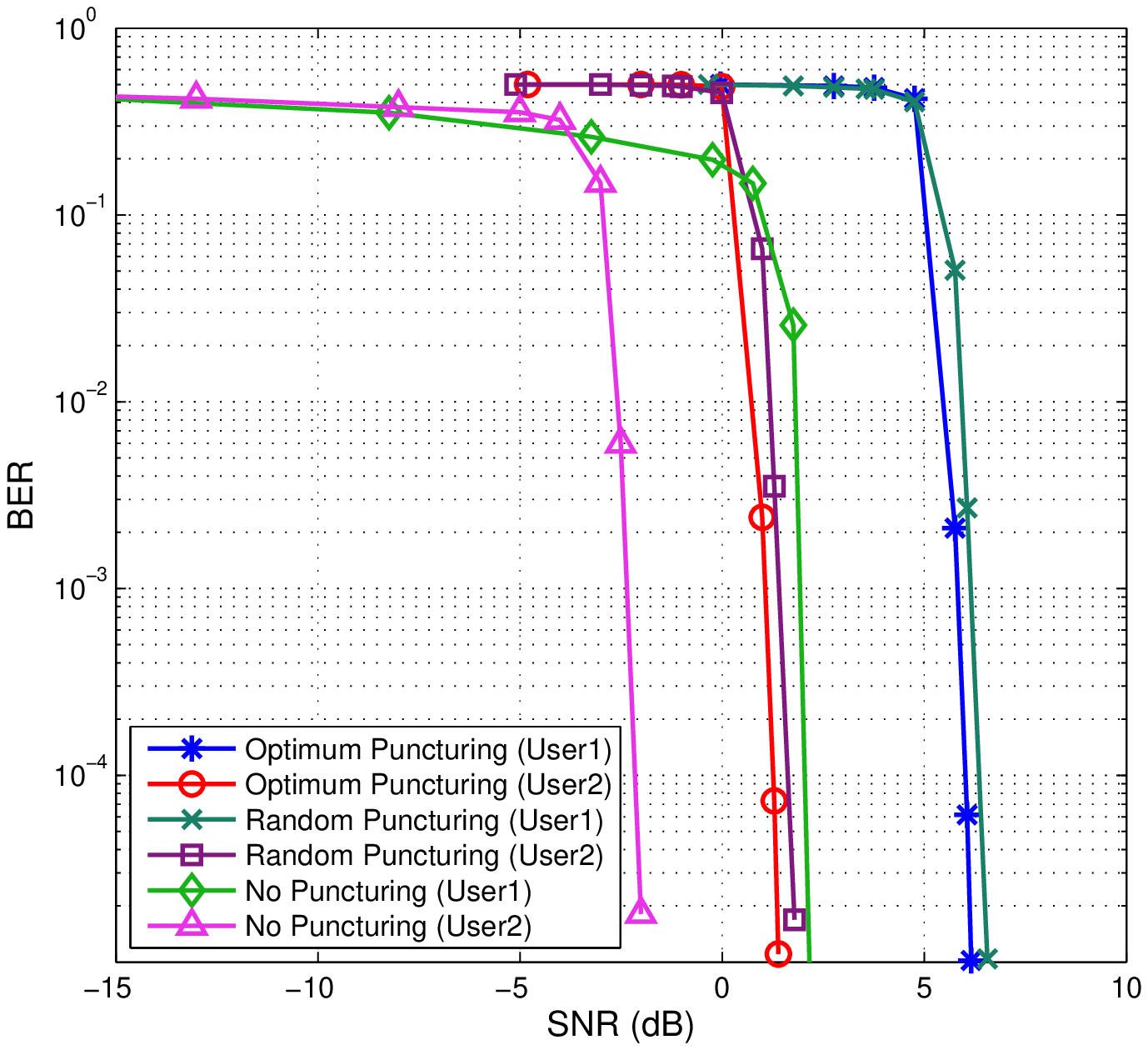}} \caption{BER versus SNR of unequal power case, with the security rates $R_{s}^{[1]}=0.4451$ and $R_{s}^{[2]}=0.2215$, for the optimized punctured, the randomly punctured and unpunctured codewords.} \label{F_unequal_BER}
\end{figure}

\begin{figure}[t]
\resizebox{\hsize}{!}{\includegraphics[scale=.3]{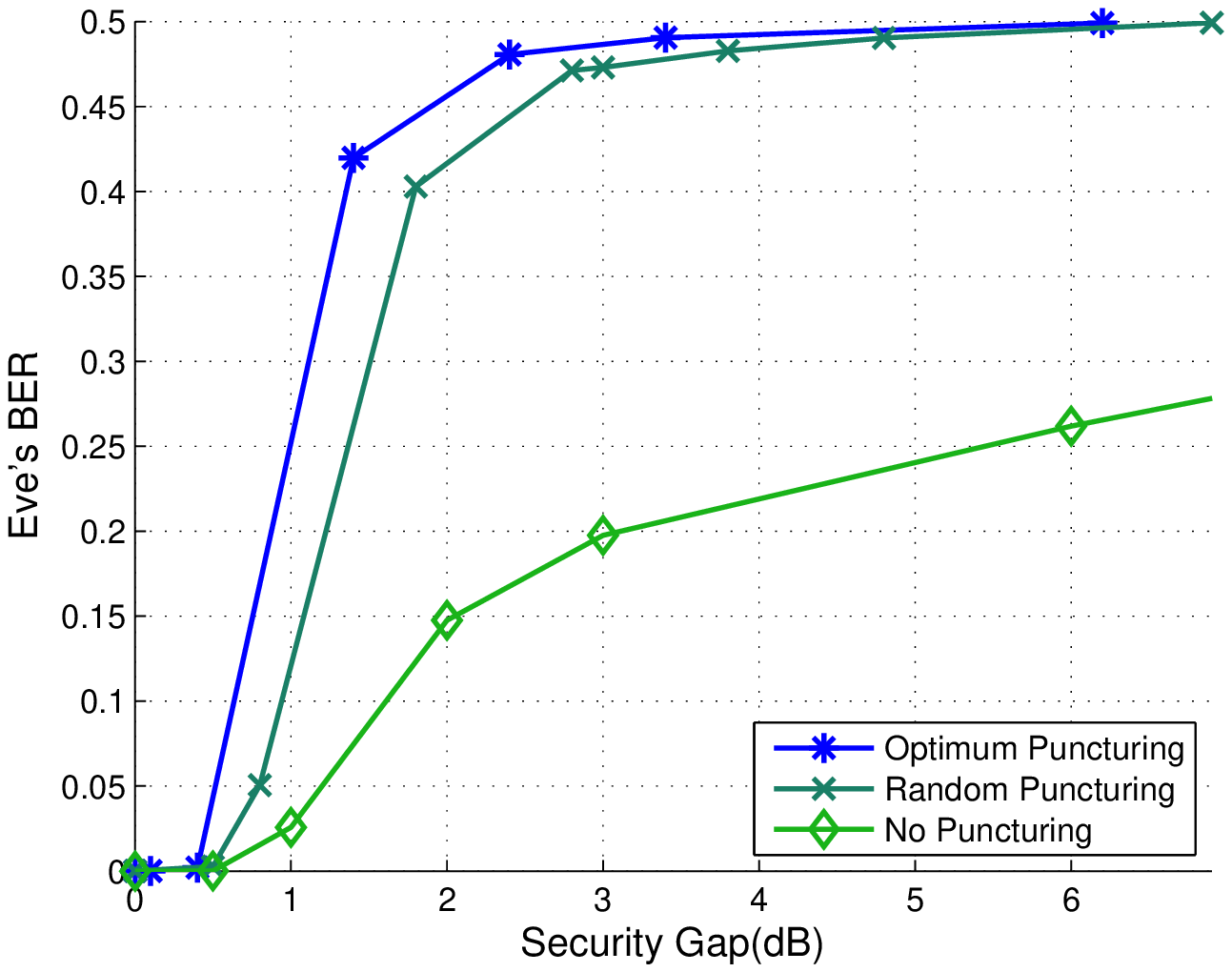}} \caption{Eve's BER versus the security gap of unequal power case, with the security rates $R_{s}^{[1]}=0.4451$ and $R_{s}^{[2]}=0.2215$, for the optimized punctured, randomly punctured and unpunctured codewords.} \label{F_unequal_gap}
\end{figure}

 The associated design rates are calculated as $R_{d}^{[1]}=0.6432$ and $R_{d}^{[2]}=0.2706$ by using (\ref{rate_R_d}) for user 1 and 2, respectively. In this case, each encoder punctures the codewords by different rates. Hence, we have to select the length of codewords according to the puncturing rates. Thus, $n^\prime_1$ and $n^\prime_2$ are considered as $14451$ and $12216$, respectively. However, the length of punctured codewords for both users equals $10^4$. According to our simulation results, the BER performance of both users in terms of SNR are depicted in Fig.~\ref{F_unequal_BER}. Furthermore, Fig.~\ref{F_unequal_gap} exhibits the BER curves corresponding to Eve versus the security gap. Using the proposed scheme for the unequal transmit power case reduces the security gap compared to the unpunctured scheme. Moreover, the optimized puncturing distribution causes extra reduction in the security gap compared to the random puncturing scheme. If Eve's BER values $P^E_{e}$ are considered to be $0.45$, $0.48$ and $0.49$, the security gap values will be $1.9$, $2.4$ and $3.4$ dB for the optimized punctured messages, and $2.5$, $3.2$ and $4.8$ dB for the random punctured messages, respectively. Additionally, they are respectively equal to $18.5$, $27.5$ and $30.5$ dB for the unpunctured messages. As seen in Fig.~\ref{F_unequal_BER}, the SNR loss is about $4$ dB and 4.4 dB for the optimized and the random puncturing methods, respectively.
\par
\begin{figure}[t]
\resizebox{\hsize}{!}{\includegraphics[scale=.3]{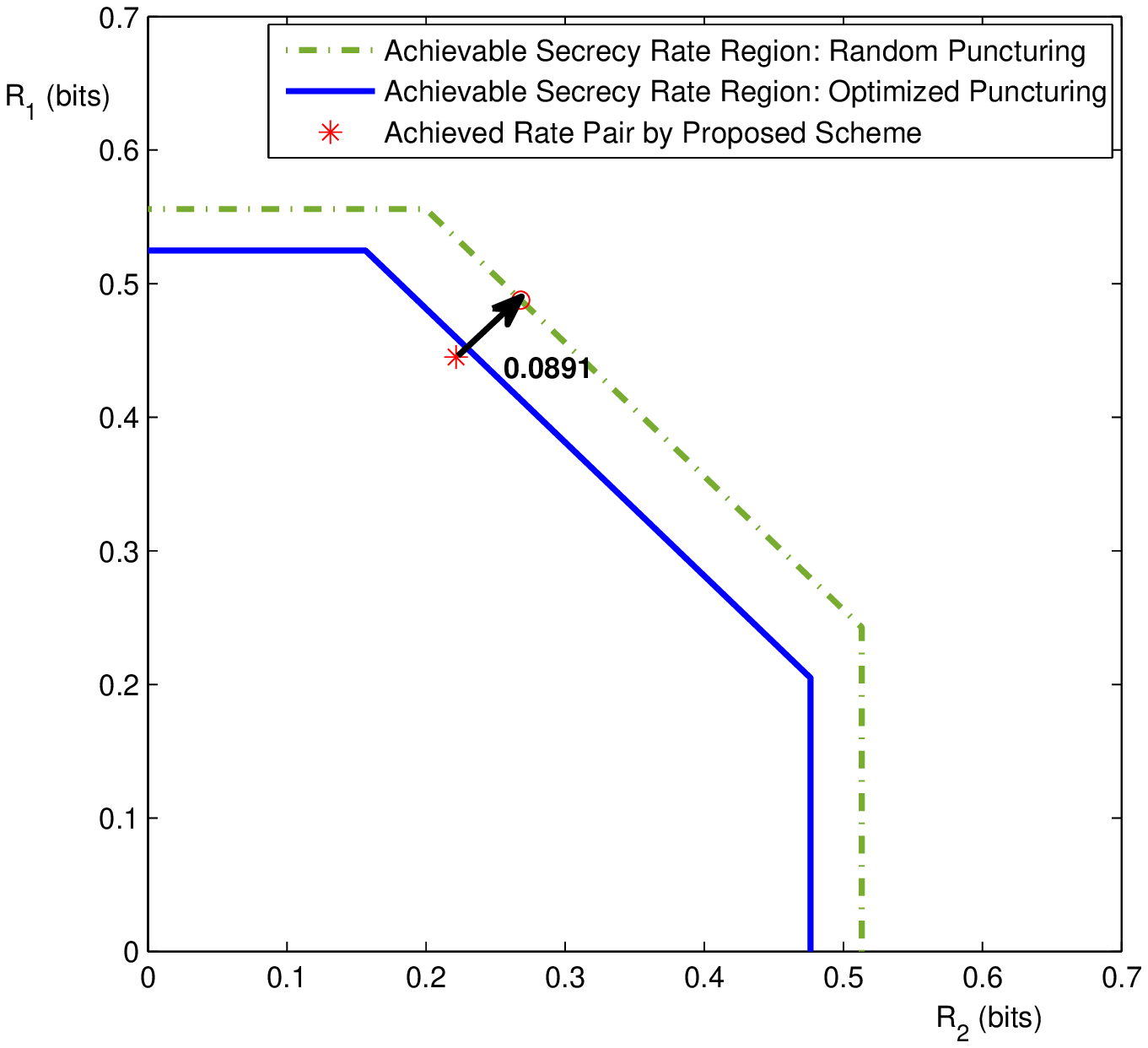}} \caption{Comparison between an achievable secrecy rate region and the obtained rate pair by applying the random and the optimized puncturing schemes for unequal transmit powers.} \label{compared_rate_region_uneq}
\end{figure}

 Similar to the equal power case, we set $P_{e}^B$ to $10^{-5}$ and $P_e^E$ to $0.5$. Hence, according to Fig~\ref{F_unequal_BER}, the noise variances of Bob's channel are obtained as $0.3631$ and $0.3311$ for the optimized and the random puncturing schemes, respectively. The noise variances of Eve's channel are also equal to $0.5136$ and $1.6218$ for the optimized and the random puncturing methods, respectively. The secrecy rate region for the mentioned cases are illustrated in Fig.~\ref{compared_rate_region_uneq}. It is apparent that the obtained sum-rate using the proposed secure encoder gets closer to the rate region by employing the optimized puncturing. In addition, the gap between the maximum sum-rate on the secrecy rate region and the achieved sum-rate is 0.0145 bits. The difference between maximum sum-rate on the secrecy rate region and the achieved sum-rate is about $0.0891$ bits for the random puncturing.

\subsection{Complexity of the secure encoder}
The main result in \cite{richardson2008modern} shows that the encoding process of LDPC codes is performed with complexity of $O(n+g_{\textrm{enc}}^2)$, where $g_{\textrm{enc}}$ is referred to the encoding gap. As shown in \cite{5740591}, the $g_{\textrm{enc}}$ is a proportional coefficient of $n$ and this coefficient is computed related to the puncturing pattern and the puncturing rate. Using two encoders in GMAC-WT, make the complexity of our proposed secure encoding scheme twice that of a single user scenario.

\section{Conclusion}
\label{sec-concul}

We have proposed an efficient scheme for sending confidential messages over the GMAC by employing LDPC codes. In this scheme, a secure encoder has been proposed by considering some random bits and puncturing the secret bits at each transmitter. To reconstruct the secret bits, we have presented a modified form of the joint decoder and its associated EXIT chart analysis. The optimization problem has been defined to maximize the puncturing rates and the corresponding secure rates. The optimized puncturing distribution has been used to achieve the reliability and the security conditions at each encoder. The security gap has been employed to determine the secrecy level of the proposed scheme. We have used random puncturing to have a comparison with the results of the optimized puncturing case. We have assessed the security level for both transmitters with equal and unequal transmit powers. Confidentiality of both users' messages has been provided by the proposed method with a fairly small security gap. The comparison between an achievable secrecy rate region and the obtained rate pair of our proposed scheme has been confirmed the advantage of the optimized puncturing case.

%\bibliographystyle{iet}
%\bibliography{IETref}

\end{document}